\documentstyle[aps,prl,multicol,epsf]{revtex}
\def\B.#1{{\bbox{#1}}}
\def\C.#1{{\cal{#1}}}
\def\BC.#1{{\bbox{\cal{#1}}}}
\def\D {\Delta}
\def\d {\delta}

\def\s {\sigma}
\def\a {\alpha}

\def\e {\epsilon}
\begin{document}
\title{Anomalous Scaling in a Model of Hydrodynamic Turbulence with
a Small Parameter}
\author{Daniela Pierotti, Victor S. L'vov, Anna Pomyalov 
and Itamar Procaccia}
\address{Department of ~~Chemical Physics, The Weizmann Institute of
Science\\Rehovot, 76100, Israel}
\maketitle

\begin{abstract}The major difficulty in developing theories for anomalous
scaling in hydrodynamic turbulence is the lack of a small parameter.
In this Letter we introduce a shell model of turbulence that exhibits
anomalous scaling with a tunable small parameter. The small parameter
$\epsilon$ represents the ratio between deterministic and random
components in the coupling between $N$ identical copies of the
turbulent field. We show that in the limit $N\to \infty$ anomalous
scaling sets in proportional to $\epsilon^4$. Moreover we give strong
evidences that the birth of anomalous scaling appears at a finite
critical $\epsilon$, being $\epsilon_c\approx 0.6$.
\end{abstract}

\begin{multicols}{2}
\narrowtext
The statistics of the small scale structure of turbulence is
characterized by ``anomalous scaling" meaning that correlation
functions and structure functions of velocity differences across a
scale $R$ exhibit a power law behavior with scaling exponents that are
not correctly predicted by dimensional analysis.  In the last few
decades there have been many attempts to compute the scaling exponents
of turbulent fields from the equations of motion. In the context of
simplified models of passive scalar advection it was discovered that
there exist natural small parameters that allow direct computations of
anomalous scaling exponents \cite{95GK,95CFKL,98GPZ}. In Navier Stokes
turbulence and also in simplified models like shell models the
progress in computing scaling exponents was slowed down by the lack of
a small parameter. It is thus worthwhile to consider models of
turbulent velocity fields in which a tunable small parameter can be
introduced and used to advantage.

We propose to introduce a small parameter via the coupling between $N$
copies of a field $u_n$ which satisfies the dynamics of the Sabra
shell model of turbulence introduced in \cite{98LPPPV}:
\begin{eqnarray}
{du_n(t)\over
dt}&=&i\Big[ak_{n+1}~u^*_{n+1}u_{n+2}+bk_{n}~u^*
_{n-1}u_{n+1}\nonumber\\
&&-c k_{n-1}~u_{n-2}u_{n-1}\Big]-\nu k_n^2~u_n+f_n(t) \ .
\label{sabra}
\end{eqnarray}
Shell models are simplified dynamical systems constructed such that
the complex number $u_n$ represents the amplitude associated with the
Fourier transform of the velocity field $\B. u(\B.r)$ with
``wave-vector" $k_n$. Rather than considering the full $\B.k$ space
and all the nonlinear interactions one allows for only one-dimensional
$k$ vectors on shells spaced such that $k_n\equiv k_0 \lambda^n$, with
$\lambda$ being the spacing parameter, and local interactions.  In
Eq. (\ref{sabra}) $\nu$ is the ``viscosity" and $f_n(t)$ a random
Gaussian force restricted to the lowest shells.  The parameters $a,~b$
and $c$ are restricted by the requirement $a+b+c=0$ which guarantees
the conservation of the ``energy" $E=\sum_{n=0}^N |u_n(t)|^2$ in the
in-viscid, unforced limit.

We now want to generalize this model to one which consist of $N$
suitably coupled copies of it.  In order to do that we need to
consider separately the equations for the real and imaginary parts of
$u_n$.  This procedure guarantees that the obtained model converges to
the original Sabra model in the limit $N\to 1$. The copies are indexed
by $i,j$ or $\ell$, and these indices take on values $-J,\dots,+J$,
$2J+1=N$. The $i$th copy of the velocity field is denoted as
$u^{[i]}_{n,\sigma}$. In this notation $\sigma=\pm 1$ refers to the
real and imaginary parts of $u_n$ respectively. Let $D^{[ij\ell]}$ be
the coupling between copies, which will be chosen later.
Equations~(\ref{sabra}) for a collection of copies are
\begin{eqnarray}
&&{du^{[i]}_{n,\sigma}\over dt}=\sum_{j\ell}D^{[ij\ell]}
\Big[A^{(\sigma)}_{\sigma'\sigma''}\Big(\gamma_{a,n+1}
u^{[j]}_{n+1,\sigma'}u^{[\ell]}_{n+2,\sigma''}\nonumber\\
&&\quad +\gamma_{b,n}
u^{[j]}_{n-1,\sigma'}u^{[\ell]}_{n+1,\sigma''}\Big)
+C^{(\sigma)}_{\sigma'\sigma''} \gamma_{c,n-1}
 u^{[j]}_{n-2,\sigma'}u^{[\ell]}_{n-1,\sigma''}\Big]\nonumber
\\&&\quad  -
\nu k_n^2~u^{[i]}_{n,\sigma}+f^{[i]}_{n,\sigma}
\ , \label{sabracopy}
\end{eqnarray}
where
\begin{equation}\label{defgammas}
\gamma_{a,n}\equiv ak_{n}\,,\quad \gamma_{b,n}\equiv b k_{n}\,,
\quad \gamma_{c,n}\equiv c  k_{n}\ .
\end{equation}
and
\begin{eqnarray}
&&\B.A^{(+1)}\equiv \left(\begin{array}{cc}1&0
\\ 0&1\end{array}\right)\ ,
\qquad \B.A^{(-1)}\equiv \left(\begin{array}
{cc}0&-1\\ 1&0\end{array}\right)\ ,
\nonumber\\
&&\B.C^{(+1)}\equiv \left(\begin{array}
{cc}-1&0\\ 0&1\end{array}\right)\ ,
\quad \B.C^{(-1)}\equiv \left(\begin{array}
{cc}0&1\\ 1&0\end{array}\right) \,,
\end{eqnarray}
Note that
$A^{(\sigma)}_{\sigma'\sigma''}=
A^{(\sigma')}_{\sigma\sigma''}\,, \qquad
A^{(\sigma)}_{\sigma'\sigma''}=C^{(\sigma')}_{\sigma''\sigma}$.

To proceed we note that the index $\ell$ is defined modulo $N$, and
introduce a Fourier transform in the ``copy" space, defining the {\it
collective} variables:
\begin{equation}
u^\alpha_{n,\sigma}={1\over \sqrt{N}}\sum_{\ell=-J}^J
u^{[\ell]}_{n,\sigma}\exp \Big ({2i\pi\alpha\ell\over N}\Big )
\ . \label{collective}
\end{equation}
Note that the index $\alpha$ is also defined modulo $N=2J+1$.  It is
convenient to consider values $\alpha$ within ``the first Brillouin
zone'' , i.e from $-J$ to $J$.  We will refer to it as the
$\alpha$-momentum. Since $u^{[i]} _{n,\sigma}$ is real,
$u^{-\alpha}_{n,\sigma}=u^{\alpha \, *}_{n,\sigma}$.
In ``$\alpha$-Fourier space'' Eqs.~(\ref{sabracopy}) read
\begin{eqnarray}
&&\!\!\!\!\!\!\!  {du^\alpha_{n,\sigma}\over dt}= \sum_{\beta,\gamma}
\Phi^{\alpha,\beta,\gamma}[\D_{\alpha,\beta+\gamma}
+\D_{\alpha+N,\beta+\gamma}  +
\D_{\alpha,\beta+\gamma+N}]\, \nonumber\\
&\times &  \Big\{A^\sigma_{\sigma'\sigma''}
\big[ \gamma_{a,n+1}
u^{\beta }_{n+1,\sigma'}u^{\gamma}_{n+2,\sigma''}
+\gamma_{b,n}  u^{\beta}_{n-1,\sigma'}u^{\gamma}_{n+1,\sigma''}
\big] \nonumber\\
&+&
C^\sigma_{\sigma'\sigma''} \gamma_{c,n-1}
 u^\beta_{n-2,\sigma'}u^{\gamma}_{n-1,\sigma''}\Big\}-
\nu k_n^2~u^\alpha_{n,\sigma}+f^\alpha_{n,\sigma}
\ . \label{sabrafourier}
\end{eqnarray}
where $\D_{\alpha,\beta}$ is the Kronecker symbol.  Observe that we
use Greek  indices for components in $\alpha$-Fourier space, and Latin
indices for copies in the copy space.  As a consequence of the
discrete translation symmetry of the copy index $[i]$
Eqs.~(\ref{sabrafourier}) conserve $\alpha$-momentum modulo $N$ at the
nonlinear vertex, as one can see explicitly in the above equation.
The coupling amplitudes $\Phi ^{\alpha ,\beta,\gamma}$ in these
equations are the Fourier transforms of the amplitudes $D^{[ij\ell]}$.

We choose the coupling amplitudes according to
\begin{equation}\label{choice}
\Phi^{\alpha,\beta,\gamma}={1\over
\sqrt{N}}\big [\epsilon+\sqrt{1-\epsilon^2}\, \Psi
^{\alpha,\beta,\gamma}\big ]\; ,
\end{equation}
where $\Psi^{\alpha,\beta,\gamma}$ are quenched random phases,
uniformly and independently distributed with zero average (cf. the
``Random Coupling Model" (RCM) for the Navier-Stokes statistics
\cite{61Kra} and the identical symmetry conditions there). 
Consequently for $\epsilon=0$ the model reduces to the RCM and
exhibits normal scaling (K41) for $N\to \infty$.  It was in fact
understood
\cite{61Kra,98Eyi} that in the limit $N\to \infty$ the
direct interaction approximation (DIA) becomes the exact
solution of the RCM.  Moreover a proper analysis of the DIA approximation
leads to normal scaling
for those systems in which sweeping effects are removed or absent
by construction like in shell models \cite{97Pie,98Eyi}.

On the other hand for $ \e=1$ the coupling coefficients in the
$\alpha$-Fourier space (\ref{choice}) are index-independent. This
corresponds to uncoupled Eqs. (\ref{sabracopy}) in the copy space,
because in this case $D^{[ij\ell]}=\d_{i,j}\d_{i,\ell}$.  Thus for $
\e=1$ we recover the original Sabra model with anomalous
scaling\cite{98LPPPV}.  Our choice of couplings (\ref{choice}) allows
an interpolation between the normal K41 scaling for $\epsilon=0$ (at
$N\to \infty$) and the full anomalous scaling of Sabra model for
$\e=1$.  A model of this type was proposed in the context of
Navier-Stokes statistics by Kraichnan in \cite{70Kra} and analyzed by
Eyink \cite{98Eyi} in terms of perturbative expansions.

\begin{figure}
 \epsfxsize=  8  truecm
 \epsfbox{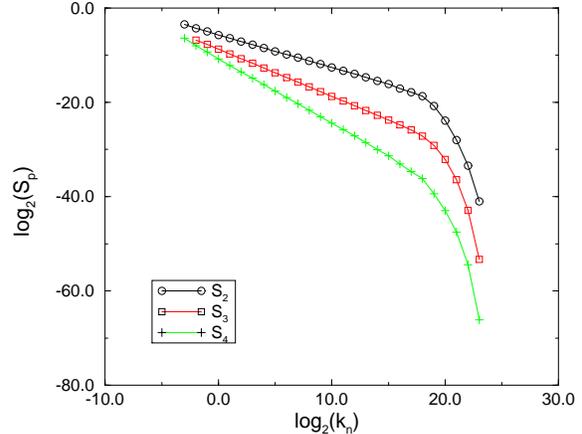}
 \caption{Log-log plot of the structure functions $S_p(k_n)$
  vs $k_n$ for p=2,3,4, $\epsilon=0.8$ and $N=25$.}
 \label{f:num09}
 \end{figure}

Our aim in this Letter is to present numerical results which show that
for small values of $\epsilon$ the model exhibits small anomalous
corrections to normal scaling and to present theoretical arguments to
rationalize the functional dependence of the anomalous corrections on
$\epsilon$.  We measured the scaling exponents of the structure
functions:
\begin{equation}
S_{2p}(k_n)\equiv \frac{1}{N}\langle \sum_{i=-J}^{J}\sum_{\sigma}
|u^{[i]}_{n,\sigma}|^{2p}\rangle \sim k_n^{-\zeta_{2p}}\; .
\end{equation}
The exponents $\zeta_{2p}$ have been calculated by a linear fit in the
two decades inertial range, see Fig.~\ref{f:num09}. The equations of
motion (\ref{sabrafourier}) with 28 shells, $a=1$, $b=c=-0.5$, were
integrated with the slaved Adams-Bashforth algorithm, viscosity
$\nu=4\times 10^{-9}$, a time-step $\Delta t=10^{-5}$.  The forcing
was subjected on the first two shells, chosen random Gaussian with
zero average and with variances such that $\s_2/\s_1=0.7$ (in order to
minimize the input of helicity\cite{98LPPPV} which leads to period two
oscillation in the structure 
\vskip -0.2cm 
%%%%%%%%%%%%%%%%%%%%%%%%%%%%
%%%%%%%%%%%%%%%%%%%%%%%%%%%%%%%
\begin{figure}
\epsfxsize=8 truecm
\epsfbox{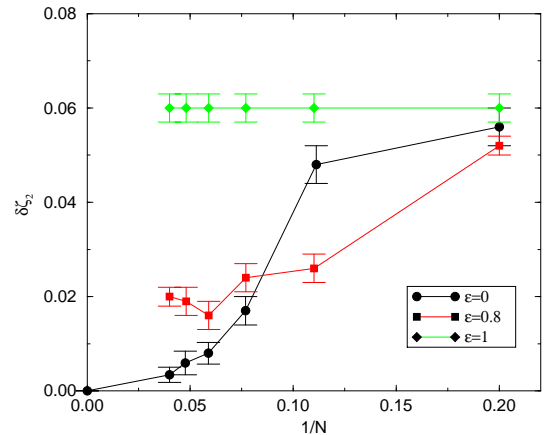}
\caption{$\delta\zeta_2=\zeta_2-2/3$ vs $1/N$ for 
$\epsilon=0$ (circles),
$\epsilon=0.8$ (squares) and $\epsilon=1$ (diamonds) for $N$ from 5 to 25.
The point at $1/N=0$ for the $\epsilon=0$ curve is the theoretical
prediction of the RCM for $N\to\infty$.}
\label{f:num08}
\end{figure}
\vskip -0.2cm 
\noindent 
functions).  Averages were taken for a
time equal to 250 eddy turnover times for the case $N=1$. The
averaging times were decreased when the number of copies increased,
taking into account the faster convergence times in these cases.  The
quality of the scaling behavior and of the fits is demonstrated in
Fig.~\ref{f:num09}.

To substantiate the birth of anomalous scaling at $\e>0$ we simulated
the model for different values of $\e$ and of $N$. We are interested
in the values of the scaling exponents for very large values of $N$
($\infty$ in theory).  In Fig.~\ref{f:num08} one can see the plot of
the value of the anomalous corrections to Kolmogorov scaling,
$\delta\zeta_2=\zeta_2-2/3$, as function of $1/N$ for $\epsilon=0.8$
together with the same curve for $\epsilon=0$ and for $\epsilon=1$ for
$N$ ranging from 5 to 25.  While for $\epsilon=0$ the corrections to
Kolmogorov scaling go to zero, for $\epsilon=0.8$ and for $\epsilon=1$
the corrections converge to a finite value which increases with
$\epsilon$.%%%%%%%%%%%%%%%%%%%%%%%%%%%% 

\begin{figure}
\epsfxsize=8 truecm
\epsfbox{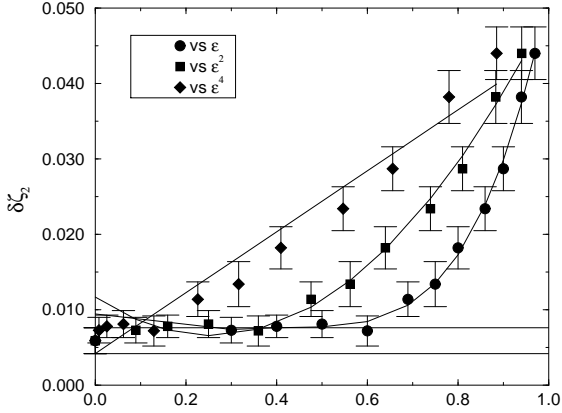}
\caption{$\delta\zeta_2=\zeta_2-2/3$ vs $\epsilon^4$
(circles), vs $\epsilon^2$ (squares) and vs $\epsilon$ (diamonds)
together with linear, quadratic and quartic fits respectively
for N=25.}
\label{f:num10}
\end{figure}
\vskip -0.6cm 
%%%%%%%%%%%%%%%%%%%%%%%%%%%%%%%
%%%%%%%%%%%%%%%%%%%%%%%%%%%%%%%
\begin{figure}
\epsfxsize=9 truecm
\epsfbox{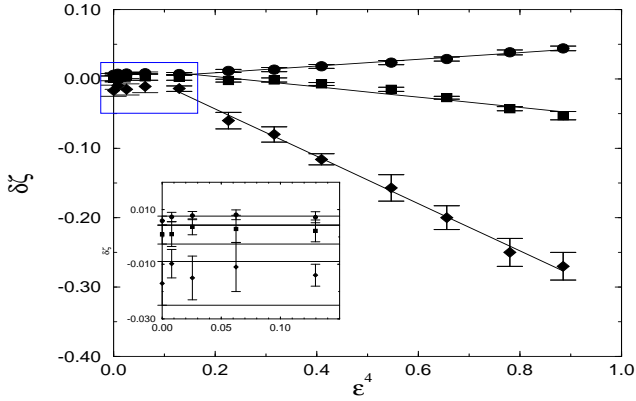}
\caption{$\delta\zeta_2=\zeta_2-2/3$ (circles), 
$\delta\zeta_4$ (squares)
and $\delta\zeta_6$ (diamonds)
vs $\epsilon^4$ for $N=25$.}
\label{f:num11}
\end{figure}
\vskip -0.2cm 
The random phases in the couplings were chosen with respect to a
uniform probability with zero-mean at the beginning of each
simulation. The rigorous procedure for quenched disorder would call
for taking averages over different runs with different choices of the
couplings. We did not do that, but rather checked that self-averaging
is already valid for $N=5$, for $\epsilon=0.8$ (small random component
in the couplings) and within our numerical precision. For $\e=0$
self-averaging occurs only for large numbers of copies.

The most interesting aspects of the numerical findings are the
dependence of the anomalies on $\epsilon$ and the question whether
anomaly appears for any $\epsilon>0$ or only above a critical value
$\epsilon_c$. The first issue is settled with sufficient clarity in
Fig.~\ref{f:num10} in which we show the behavior of $\delta\zeta_2$
(for N=25) as a function of $\epsilon$, $\epsilon^2$ and $\epsilon^4$
with the respective linear, quadratic and quartic fits.  One can see
clearly that the anomalous corrections $\delta\zeta_2$ go to zero like
$\epsilon^4$. The same behavior is exhibited by $\delta\zeta_4$ and
$\delta\zeta_6$ as one can see in Fig ~\ref{f:num11}.
\begin{figure}
\epsfxsize=8 truecm
\epsfbox{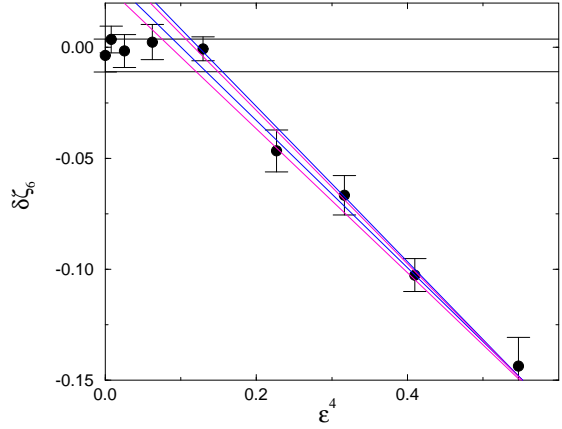}
\vskip -0.4cm 
\caption{$\delta\zeta_6=\zeta_6-2$
vs $\epsilon^4$ for $N=25$ with the maximal and minimal slope
lines obtained for the linear fits in the two ranges
$\epsilon^4\in[0.13,0.88]$ and $\epsilon^4\in[0.23,0.88]$ .
All the values of the anomalous corrections have been shifted by
the average of the first five points.}
\label{f:num12}
\end{figure}

\vskip -0.4cm 
The existence of a critical value $\epsilon_c$ is harder to settle. In
the following we show arguments in favor of the existence of a finite
$\epsilon_c$.  First one should notice in the magnification in
Fig.~\ref{f:num11} that at $\epsilon=0$ we do not get the K41 values
$\delta\zeta_n=0$ as theoretically predicted. This is a result of the
finiteness of the number of copies ($N=25$ in Fig.~\ref{f:num11}).

To establish the existence of a finite $\epsilon_c$ we proceeded as
follows: i) To take this into account the finite size effects we
subtracted the $\epsilon=0$ value of $\delta\zeta_n$ from all the
values of $\zeta_n$.  ii) We calculated the maximal and minimal slope
line, that is the best fit slope plus and minus the standard
deviation, for $\delta\zeta_n$ vs $\epsilon^4$ in two different
ranges: $\epsilon^4\in[0.13,0.88]$ and $\epsilon^4\in[0.23,0.88]$:
that is, we fitted with the 2 parameter function
$f(\epsilon^4)=a_n+b_n\epsilon^4$.  iii) We fitted the plot of
$\delta\zeta_n$ vs $\epsilon$ with the 4 parameter function
$g(\epsilon)=a_n\epsilon+b_n\epsilon^2 +c_n
\epsilon^3+d_n\epsilon^4$
in the range $\epsilon\in[0,0.97]$.Note that with the ``calibration''
of the zero that we performed $\delta\zeta_n=0$ for $\epsilon=0$.  The
$\chi^2$ test is much better for the first procedure than for the
second, where the minimal $\chi^2$ is tripled, although in principle
it should be easier to fit a function with a larger number of
parameters.

\begin{table}
\begin{tabular} {||c||c|c|c||}
$\delta\zeta_n$ & range of $\epsilon^4$ 
& slope$\times 10^3$ & $\epsilon_c$
\\ \hline\hline
$\delta\zeta_2$ & [0.13,0.88] &48$\pm$ 3  & 0.60 $\pm$ 0.07
\\ \hline
$\delta\zeta_2$ & [0.23,0.88] & 50$\pm$ 3  & 0.62 $\pm$ 0.06
\\ \hline\hline
$\delta\zeta_4$ & [0.13,0.88] & $-74 \pm 8$ & 0.69 $\pm$ 0.07
\\ \hline
$\delta\zeta_4$ & [0.23,0.88] & $-82 \pm 9$ & 0.71 $\pm$ 0.06
\\ \hline\hline
$\delta\zeta_6$ & [0.13,0.88] & $-341 \pm 8$  & 0.59 $\pm$ 0.04
\\ \hline
$\delta\zeta_6$ & [0.23,0.88] & $-335 \pm 6$ & 0.57 $\pm$ 0.05
\\
\end{tabular}
\vskip 0.3cm
\caption{ Results of the linear fits on different ranges
for different scaling exponents. The values of $\epsilon_c$
have been calculated as explained in the body of the text.}
\label{tab1}
\end{table}
 
 We interpret this result as an evidence for the existence
of a finite $\epsilon_c$.

In order to have a better estimate of the ``zero'' level of
$\delta\zeta_n$'s, and so a better estimate of $\epsilon_c$, instead
of subtracting the value of the anomalous correction at $\epsilon=0$
we subtracted the average value of $\delta\zeta_n$ calculated by using
the first 5 points belonging to the flat region.  The value of
$\epsilon_c$ with its error has been found by looking at the
intersections of the minimal and maximal slope lines with the zero
line: the line of the value of $\delta\zeta_n$ for $\epsilon=0$ with
its error (see Fig.~\ref{f:num12}).  Table~\ref{tab1} exhibits the
resulting values for the various exponents and their linear fits. Note
that all the values of $\epsilon_c$ coincide within the error bars.

To understand the linear dependence on $\epsilon^4$ we turn to the
exact equations that are satisfied by the $n$-order correlation functions,
which are defined as suitable averages over time and over all the replica.
The second and third order correlation functions are defined as:
\begin{eqnarray}
&& F_2(k_n;t-t')\equiv \frac{1}{N}\sum_{\alpha,\sigma}
\langle u_{n,\sigma}^\alpha(t) u_{n,\sigma}^{\a*}(t')\rangle
\, , \label{F2}\\
&&F_3(k_n;t,t',t'') \equiv \frac{1}{N}\sum_{\alpha,\alpha',\alpha''}
\sum_{\sigma,\sigma',\sigma''}
\Phi^{\alpha'',\alpha',\alpha}\\
&& \big[\D_{\a+\a',\a''}+\D_{\a+\a',\a''+N}
+\D_{\a+\a'+N,\a''} \big]  \nonumber\\
&&\langle
u_{n-1,\sigma'}^{\alpha}(t) u_{n,\sigma'}^{\a'}(t')
u_{n+1,\sigma''}^{\alpha''*}(t'') \rangle \ .\nonumber
\label{tab1}
\end{eqnarray}
and similarly for higher order correlation functions and for the
response functions (for details see \cite{98LPPP}).

The equations of motion for these objects can be written down
explicitly. For example,
\FL
\begin{eqnarray}\label{hierF3}
&&{\partial \over \partial t}F_2(k_n,t)
= \gamma_{a,n+1}F_3(k_{n+1};0,t,t)
+ \gamma_{b,n}F_3(k_{n};t,0,t)\nonumber\\
&&+ \gamma_{c,n-1} F_3(k_{n-1};t,t,0)\, ,\\
&&{\partial \over \partial t_1}
F_3(k_n,t_1,t_2,t_3)=
\gamma_{a,n}
F_4(k_{n},k_{n+1},k_{n},k_{n+1};t_1,t_1,t_2,t_3)\nonumber\\
&&+ \gamma_{b,n-1}
F_4 (k_{n-2},k_{n},k_{n},k_{n+1};t_1,t_1,t_2,t_3)\nonumber\\
&&+ \gamma_{c,n-2}
F_4 (k_{n-3},k_{n-2},k_{n},k_{n+1};t_1,t_1,t_2,t_3)\,,  \label{hierF41}
\end{eqnarray}
In order to close such equations and to attempt to solve them,
one needs to express a higher order correlation function in
terms of lower order statistical objects. In \cite{98LPPP} it was
shown that this can be done in the present context in a controlled
fashion. In other words, it is possible to express, say,
$F_4$ in terms of $F_2$, $F_3$ and second and third order
response functions. In doing so one of course leaves out information
about $F_4$ that cannot be possibly represented in terms of lower
order objects. Yet, the main result of the analysis of \cite{98LPPP}
is that the neglected terms in this procedure are of $O(\epsilon^6)$ whereas
the retained terms are of $O(1)$ and of $O(\epsilon^4)$!
Since we understand that for $\epsilon=0$ the anomaly must vanish,
we expect the anomalies to be proportional to $\epsilon^4$.
We interpret therefore the numerical results shown in Fig ~\ref{f:num10}
and Fig ~\ref{f:num11} as an excellent confirmation of
this theoretical expectation.

To summarize: we introduced a shell model of hydrodynamic turbulence
in which a small tunable parameter $\epsilon$ exists. We know from
theoretical and numerical results that the model exhibits normal
scaling for $\epsilon=0$. The most important results of this letter
are the existence of a finite $\epsilon_c$ at which anomalous scaling
appears and the behavior as $\epsilon^4$ of the anomalous corrections
for $\epsilon>\epsilon_c$.

\acknowledgments This work has been supported in
part by the European Commission under the Training and Mobility of
Researchers program, The German-Israeli Foundation, the Israel Science
Foundation administered by the Israel Academy of Sciences, and the
Naftali and Anna Backenroth-Bronicki Fund for Research in Chaos and
Complexity.

%%%%%%%%%%%%%%%%%%%%%%

\end{multicols}


\begin{references}
\bibitem{95GK}
K. Gawedczki and A. Kupiainen, Phys. Rev. Lett. {\bf 75}, 3834 (1995)

\bibitem{95CFKL}
M. Chertkov, G. Falkovich, I. Kolokolov and V. Lebedev, Phys. Rev. E {\bf 52},
4924 (1995)

\bibitem{98GPZ}
O. Gat, I. Procaccia and R. Zeitak, Phys. Rev. Lett. {\bf 80}, 5536 (1998).

\bibitem{98LPPPV}
V.S. L'vov, E. Podivilov, A. Pomyalov, I. Procaccia and D. Vandembrouq,
Phys. Rev. E, {\bf 58} 1811 (1998).

\bibitem{61Kra}
R. H. Kraichnan, J. Math. Phys. {\bf 2}, 124 (1961).

\bibitem{98Eyi}
G. L. Eyink, {\em Random Coupling Model and Self-Consistent
$\epsilon$-Expansion Method}, (unpublished).

\bibitem{97Pie}
D. Pierotti, Europhys. Lett. {\bf 37}, 323 (1997).

\bibitem{70Kra} R. H. Kraichnan, J. Fluid Mech. {\bf 41} 189 (1970).

\bibitem{98LPPP}
V.S. L'vov, D. Pierotti, A. Pomyalov, I. Procaccia,
submitted to Phys. Fluids, special issue in Honour
of R.H. Kraichnan (1998).

\end{references}
\end{document}